\documentclass{aa}

\usepackage{txfonts}

\usepackage{graphicx}

\usepackage{natbib}
\bibpunct{(}{)}{;}{a}{}{,} 

\newcommand{\ten}[2]{#1\times 10^{#2}}
\newcommand{\teff}{$T_\mathrm{eff}$}
\newcommand{\logg}{$\log g$}
\newcommand{\nh}{$N_\mathrm{HI}$}

\newcommand{\xh}{$x_\mathrm{H}$}
\newcommand{\xhe}{$x_\mathrm{He}$}

\newcommand{\anomi}{$A^\mathrm{0}$}

\newcommand{\chisq}{$\chi^2$}
\newcommand{\chisqred}{$\chi^2_\nu$}

\newcommand{\orx}{\mbox{\object{RX\,J185635$-$3754}}}
\newcommand{\rxj}{\mbox{{RX\,J1856}}}
\newcommand{\ohz}{\mbox{\object{HZ43\,A}}}

\newcommand{\osi}{\mbox{\object{Sirius\,B}}}

\newcommand{\cmcub}{cm$^{-3}$}

\newcommand{\ergsa}{erg\,cm$^{-2}$s$^{-1}$\AA$^{-1}$}
\newcommand{\photsa}{photons~cm$^{-2}$s$^{-1}$\AA$^{-1}$}
\newcommand{\ctsa}{cts~s$^{-1}$\AA$^{-1}$}

\newcommand{\abc}{$a(C)$}

\newcommand{\rosat}{\textit{ROSAT}}

\newcommand{\chandra}{\textit{Chandra}}
\newcommand{\euve}{\textit{EUVE}}

\begin{document}

\title{Establishing \ohz, \osi, and \orx\ as soft X-ray standards: a
cross-calibration between the Chandra LETG+HRC-S, the EUVE spectrometer,
and the ROSAT PSPC\thanks{Based on observations collected with the
\chandra\ Observatory, the \textit{Hubble} Space Telescope, the \euve\
satellite, and \rosat.}}

\author{K.~Beuermann\inst{1} \and
  V.~Burwitz\inst{2} \and
  T.~Rauch\inst{3}}
  
  
  \institute{Institut f\"ur Astrophysik, Friedrich-Hund-Platz~1,
    D-37077~G\"ottingen, Germany,
    \email{beuermann@astro.physik.uni-goettingen.de} \and MPI f\"ur
    Extraterrestrische Physik, D-85748 Garching, Germany,
    \email{burwitz@mpe.mpg.de} \and Institut f\"ur Astronomie und
    Astrophysik, Sand~1, D-72076 T\"ubingen, Germany,
    \email{rauch@astro.uni-tuebingen.de}}

  \date{Received April 24, 2006\ / Accepted June 29, 2006}
  
\abstract { 
The absolute calibration of space-borne instruments in the soft X-ray
regime rests strongly on model spectra of hot white dwarfs.}
{ 
We analyze the \chandra\ LETG+HRC observations of the white dwarfs
\ohz\ and \osi\ and of the neutron star \orx\ with the aim of
resolving current uncertainties in the soft X-ray spectral fluxes and
photospheric parameters of the three stars. We apply the derived
photon spectra to a cross-calibration of the LETG+HRC-S with the
short-wavelength \euve\ spectrometer and the \rosat\ PSPC.}
{ 
We tie \ohz\ to the flux of \rxj\ in the 44--48\AA\ range and perform
a simultaneous least squares fit to the LETG+HRC spectra of the three
stars.  This allows us to determine an internally consistent set of
spectral energy distributions and an empirically derived
wavelength-dependent correction to the LETG+HRC-S effective area. We
employ NLTE model atmospheres calculated with \textit{TMAP} for the
white dwarfs and a two-blackbody model for \rxj, tied to the
respective optical fluxes.}
{ 
The two-blackbody model for \rxj\ features a hot spot on a cooler star
and yields k$T_\mathrm{spot}=62.8\pm0.4$\,eV and
k$T_\mathrm{star}=32.3\pm0.7$\,eV with a stellar angular radius as
seen from infinity of $0.1371\pm0.0010$\,km\,pc$^{-1}$. For \ohz, our
fit yields \teff=51126$\pm 660$\,K and \logg=7.90$\pm0.080$ (cgs) with
anti-correlated errors (1-$\sigma$) which include not only the
statistical but also the systematic uncertainties of the fit. HZ43AB
displays a previously detected bremsstrahlung component with a
temperature k$T\simeq 0.6$\,keV.  For \osi, we find \teff=24923$\pm
115$\,K for fixed \logg=8.6.  The calibration of the short-wavelength
\euve\ spectrometer differs from that of the LETG+HRC-S by $15\pm
7$\%. The \rosat\ PSPC is found to be correctly calibrated within a
few percent and reports of a major miscalibration are unfounded. }
{ 
We have obtained improved parameters for \orx, \ohz, and \osi\ which
fit the observations from the optical to the soft X-ray regime. Our
approach allows us to quote their absolute spectral fluxes at selected
wavelengths which may aid the calibration of other space-borne
instruments.}  
   
\titlerunning{Soft X-ray standard stars \ohz, \osi, and \orx}
\authorrunning{K.~Beuermann et al.}

\maketitle


\section{Introduction}

Hot white dwarfs with pure hydrogen atmospheres are well established
standard stars in the optical and ultraviolet
\citep[e.g. ][]{bohlinetal01} and have played a decisive role in the
calibration of satellite instruments in all spectral regimes from the
optical to the extreme ultraviolet and soft X-rays. The FOS and STIS
on the \textit{Hubble} Space Telescope \citep{bohlin00}, the Extreme
Ultraviolet Explorer (\textit{EUVE}) \citep{boydetal94}, and the Low
Energy Transmission Grating Spectrometer on board \textit{Chandra}
(LETG+HRC-S) with the High Resolution Camera as detector
\citep{brinkmanetal00,peaseetal00} are prominent examples. The latter
two instruments were calibrated on the ground, but their calibration
was partly relinquished in favor of an in-flight calibration using hot
white dwarfs. The Position Sensitive Proportional Counter (PSPC) on
board the R\"ontgen Satellite (\rosat), on the other hand, was
accurately calibrated on the ground
\citep{truemper82,pfeffermannetal87} and not recalibrated in
flight. Based on observed PSPC spectra of white dwarfs, however,
\citet{napiwotzkietal93,jordanetal94, wolffetal95,wolffetal96} claimed
the PCPC calibration at soft X-ray energies to be incorrect by a
factor of about two. Given this important role of white dwarfs it is
deplorable that their soft X-ray fluxes are much less well documented
than their optical and ultraviolet fluxes. Given, furthermore, the
large body of scientific data of all types of objects potentially
affected, the resolution of such open calibration issues is of general
interest.

Our initial aim was to investigate the PSPC calibration which
requires, however, to take recourse to spectrally higher resolved data
as provided by the \euve\ or the \chandra\ LETG+HRC
spectrometers. Since the effective area of the latter is itself
uncertain by as much as 25\% at intermediate wavelengths around 60\AA\
\citep[e.g.][]{peaseetal03}, we found it necessary to perform a more
detailed cross-calibration of the three instruments. As a result, we
are able to determine accurate soft X-ray fluxes of the hot white
dwarfs \ohz\ and \osi\ and of the neutron star \orx\ (henceforth \rxj)
and thereby establish them as standards in the soft X-ray
regime\footnote{We consider neither quasars and BL\,Lacs because of
their inherent variability nor supernova remnants because of their
extended nature.}. Although one of the two white dwarfs (\ohz)
is a well known standard for wavelengths $\lambda > 912$\AA\
\citep{bohlinetal01} and the optical flux of the other (\osi) is
reasonably well measured \citep{barstowetal05}, the extrapolation of
the optically adjusted model spectra into the soft X-ray regime
suffers from the wagging-tail problem. We solve this problem by tying
the soft X-ray tail of the spectrum of \ohz\ to that of \rxj.  This
adds a second fix to the spectrum of \ohz\ besides that in the
optical and effectively constrains its spectral shape.

The organization of the paper is as follows. In Sect.~2, we describe
the analysis of the LETGS and PSPC data and list the EUVE archive data
used. Sect.~3 is devoted to the description of our theoretical white
dwarf spectra and summarizes the available information on \teff\ and
\logg\ of \ohz. Our principal results are provided in Sect.~4. It
starts with information on the optical fluxes to which our model
spectra are normalized (Sect.~4.1). In Sect.~4.2, we derive an
accurate spectrum of \rxj\ that serves as the short-wavelength fix for
\ohz. In this context, we discuss interstellar absorption near the
carbon K-edge at 43.7\AA\ to some detail. The essence of the paper is
then presented in Sect.~4.4, the simultaneous fit of \rxj, \ohz, and
\osi\ that results in an internally consistent set of photon
spectra and simultaneously in the wavelength-dependent correction
functions to the \mbox{LETG+HRC-S} effective areas at wavelengths
$\lambda \ge 44$\AA. Here, we discuss also the systematic errors of
our approach. The stellar parameters implied by our fit are presented
and discussed in Sect.~4.5. Sect.~4.6 compares our LETGS calibration
to that of the \euve\ short wavelength spectrometer and Sect.~4.7
finally demonstrates the accurate calibration of the \rosat\ PSPC. The
conclusions are summarized in Sect.~5.

\section{Observations and Data Analysis}

We have extracted all observations of \ohz, \osi, and \rxj\ taken with
the \chandra\ LETG+HRC-S, the \euve\ spectrometer, and the \rosat\
PSPC from the respective archives.  Table~1 contains a log of the
observations. 

\begin{figure}[t]
\vspace*{0.4mm}
\includegraphics[width=8.8cm]{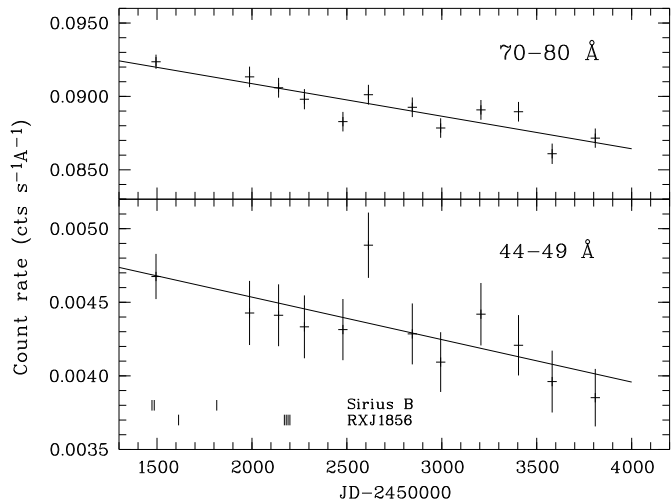}
\caption{Long-term variation of the integrated level 2.0 LETG+HRC
count rates of \ohz\ for two spectral intervals. The two sets of tick
marks indicate the times when the LETG+HRC spectra of \osi\ and \rxj\
were taken.}
\label{fig:timevar}
\end{figure}

\subsection{\chandra\ LETG+HRC spectra}

The spectra were extracted from the reprocessed level 2.0 event
files. The corresponding total exposure times are listed in
Table~\ref{tab:obslog}. The extraction region recommended in the
Chandra Proposers' Observation Guide was used and the background
determined from wide regions below and above the source extraction
area. Further details are given by \citet{burwitzetal03}. Since the
background is variable, we investigated the influence of rigorously
discarding intervals of high background on the resulting count rate
spectra. Our tentative elimination scheme (in short referred to as
level 3) reduces the exposure times of \rxj, \ohz, and \osi\ to
66\%, 84\%, and 61\% of the level 2.0 accepted times, respectively,
and is more restrictive than the condition which \citet{drakeetal02}
imposed on the accepted telemetry rate. Nevertheless, the
wavelength-integrated count rates increase by only 0.8\%, 0.1\%, and
6.8\%, respectively. These values measure the additional dead-time
resulting from the time intervals of enhanced background. We opt to
accept the level 2.0 spectra with the small dead-time corrections
applied to them as a multiplicative factor. Within their statistical
errors, the level 3 and the dead-time corrected level 2.0 spectra are
indistinguishable frome one another.

The long series of spectra taken of \ohz\ between 1999 and 2006 lends
itself to an analysis of the long-term behavior of the HRC
sensitivity. The zeroth order count rate of \ohz\ decreases slowly
with time at a rate of $0.98\pm0.15$\% per year or with a time scale
of about 100 years. This decrease seems to depend on wavelength, with
a low of the decay rate of 0.5\% per year near 110\AA, rates of
$0.9\pm 0.2$\% for 70--80\AA, and a more uncertain $2.4\pm 0.8$\% for
44--49\AA\ (Fig.~\ref{fig:timevar}). For \osi\ and \rxj, the time
bases are much more restricted as indicated by the tick marks in
Fig.~\ref{fig:timevar}. The decay rates of their zeroth order count
rates are $\sim0.7$\% and close to zero, respectively. The spatial
and temporal gain changes of the HRC-S are a known effect and a study
for times before JD2452900 has been presented by Pease \& Drake (2003).
We do not attempt to correct for these changes, but rather chose a
pragmatic approach. We place a simultaneous spectral fit on a secure
basis by selecting a subset of the \ohz\ exposures with an epoch
nearly equal to that of the \rxj\ observations, i.e., the
exposure-weighted mean Julian day JD\,2452130. We use the \ohz\
exposures 00059 through 03677 which comprise 157239 s of exposure with
an epoch different from that of \rxj\ by only 38 days. A similar
selection is not possible for \osi\ which is centered 508 days
earlier. We do not correct for the possible 1\% effect.

\begin{table}[t]
\caption{Journal of observations extracted from the \chandra, \euve,
and \rosat\ archives for the present analysis. Only the total exposure
times are given, except for the \rosat\ PSPC where the experimental
setups differ and the individual exposure times are given.}
\label{tab:obslog}
\begin{tabular}{lllr} 
\hline \hline \noalign{\smallskip}
Object &  Obs.~Identif. & Year &     Exp. (ks)\\ 
             
\noalign{\smallskip} \hline
\noalign{\smallskip}
\multicolumn{4}{l}{\textit{a) Chandra LETG+HRC-Spectrometer}} \\[0.5ex]
\ohz   & 00059, 01011, 01012, & 1999, 2001 & 254.8\\
        & 02584, 02585, 03676, & 2002       &      \\
        & 03677, 05042, 05044, & 2003, 2004 &      \\
        & 05957, 05959, 06473  & 2005, 2006 &      \\[0.5ex]
\osi & 01421, 01452, 01459  & 1999, 2000 &  64.1\\[0.5ex]
\rxj    & 00113, 03380, 03381, & 2000, 2001 & 501.9\\
        & 03382, 03399         &            &      \\[0.5ex]

\multicolumn{4}{l}{\textit{b) EUVE-Spectrometer}}\\[0.5ex]
\ohz   & 1904N, 1620N         & 1994, 1995 & 130.0\\
        & 1049N, 0030N         & 1996, 1997 &      \\[0.5ex]
\osi & 0147N, 0039N         & 1993, 1996 & 246.0\\[0.5ex]

\multicolumn{4}{l}{\textit{c) ROSAT PSPC}}\\[0.5ex]
\ohz   & rp100308             & 1990       &  21.5\\
        & rf200418             & 1991       &  21.6\\[0.5ex]
        & rp141916/17          & 1992       &   6.9\\[0.5ex]
\osi & rf200422             & 1991       &   3.2\\[0.5ex]
\rxj    & rp200497             & 1992       &   6.3\\[0.5ex]
\noalign{\smallskip} \hline      
			         
\end{tabular}
\end{table}

We start using the first-order effective areas of the LETG+HRC-S in
the positive and negative dispersion directions as recommended in the
November 2004 release of the \chandra\ X-ray
Center\footnote{http://cxc.harvard.edu/cal/Letg $\rightarrow$
LETG/HRC-S Effective Area (updated November 2004).} and refer to them
as the nominal areas \anomi. Throughout this paper, we use the
efficiencies $\varepsilon_\mathrm{k}$ of orders k~=~2 to 6 relative to
the first order as given in the same release. The areas \anomi\ are
reliable to better than 15\% shortwards of the carbon K-edge at
$\lambda=43.7$\AA, but may need corrections possibly as large as 25\%
in the 60--80\AA\ regime \citep{peaseetal03}. We distinguish,
therefore, \anomi\ from $A=\alpha$\anomi, where $\alpha(\lambda)$ is a
wavelength dependent adjustment factor which equals unity shortwards
of the carbon K-edge and is allowed to deviate from unity longwards of
it. Our correction function $\alpha$ derived below is strictly valid
only for the epoch JD\,2452130 and may differ for other epochs during
the lifetime of the HRC at the percent level.

The calibration of the LETG+HRC-S at long wavelengths is based, in
part, on \osi\ and \ohz\ \citep{brinkmanetal00,peaseetal00}, with
a certain preference given to \osi\ because of its better defined
\logg\ \citep{barstowetal05,holbergetal98}. The drawback is the lack
of a useful signal from \osi\ at $\lambda < 60$\AA, while \ohz\
can be detected at least down to $\lambda =44$\AA. Since we wish to
connect the spectrum of our white dwarf calibrator to the harder
spectrum of \rxj, we give preference to \ohz. This has the added
advantage that \rxj\ and \ohz\ are both observed on-axis while the
Sirius pointings are slightly off axis.The latter may affect the
simultaneous fit if the HRC efficiency differs spatially (in addition
to temporally) for a given wavelength. We neglect this possible
effect.

We bin the LETG+HRC spectra to 0.5\AA\ in wavelength which is an
adequate compromise between spectral resolution and statistical
significance of individual data points in the less well exposed
wavelength regions. The detector gaps for the observations of \rxj\
and \ohz\ are conservatively defined as 49--58\AA\ and
\mbox{58--69\AA} for the negative and positive dispersion directions,
respectively. For \osi, the gaps overlap and statistically significant
data of exist only for $\lambda \ge 61$\AA.

\subsection{\euve\ spectra}

We obtained six night time \euve\ spectra of \ohz\ and two of
\osi. The archived photon spectra were used and we refrained from a
re-extraction of the data. The archived spectra are based on corrected
effective areas as described by \citet{boydetal94}. Two spectra of
\ohz\ (9302180059, 9403252311) were discarded because of excessive
noise and low flux, respectively. The remaining four are listed in
Table~\ref{tab:obslog}. They differ in their general flux levels by
$\pm 5$\%, but no systematic time variation is seen. We did not use
the very faint and ill-determined \euve\ spectrum of \rxj\ discussed
by \citet{ponsetal02}.

\subsection{\rosat\ PSPC spectra}

All \rosat\ spectra of \ohz, \osi, and \rxj\ taken with the Position
Sensitive Proportional Counter (PSPC) were extracted from the \rosat\
archive and re-analyzed. Since the three sources are very bright, we
used a generous extraction radius of 5\,arcmin radius which was only
reduced for the determination of the high energy bremsstrahlung tail
of \ohz\ \citep{odwyeretal03}. The background was taken, if possible,
from a surrounding ring free of additional faint sources. We list in
Table~\ref{tab:obslog} the subset of observations discussed in detail
here, one each from the 1990 Program Verification (PV) phase, the 1991
Announcement of Opportunity 1 (AO-1), and the 1992 AO-2. The PV
observations were taken with PSPC-C and the AO-1 observations with
PSPC-B before the reduction of the high voltage on October 11,
1991. The detector response matrix for both data sets is DRMPSPC-AO1
\citep{brieletal95}. For the AO-2 observations after the voltage
change , the appropriate matrix is DRMPSPC. In order to allow a
straightforward comparison with the previous analyses of \citet[][and
references therein]{wolffetal96}, we do not apply any adjustments to
the matrices and refrain from corrections for the known (small)
temporal and spatial gain shifts of the PSPC
\citep{prietoetal96,snowdenetal01}. For \osi\ and \rxj\ only one PSPC
observation each was performed, of which the former is in AO-1 and the
latter in AO-2. The observations with identification numbers starting
with rp were taken in the open configuration, the ones starting with
rf include the boron filter which reduces the count rate for the very
soft white dwarfs by about a factor of eight as a result of the added
absorption and the narrower width of the response in photon energy.

All integrated PSPC count rates quoted in Sect.~4.7 below refer to
energy channels 11--60 (of 256). We note already here that our count
rates are lower than those presented by \citet{wolffetal96} who seem
to have included channels below the recommended lower limit of channel
11.

\begin{figure}[t]
\vspace*{0.4mm}
\includegraphics[width=8.8cm]{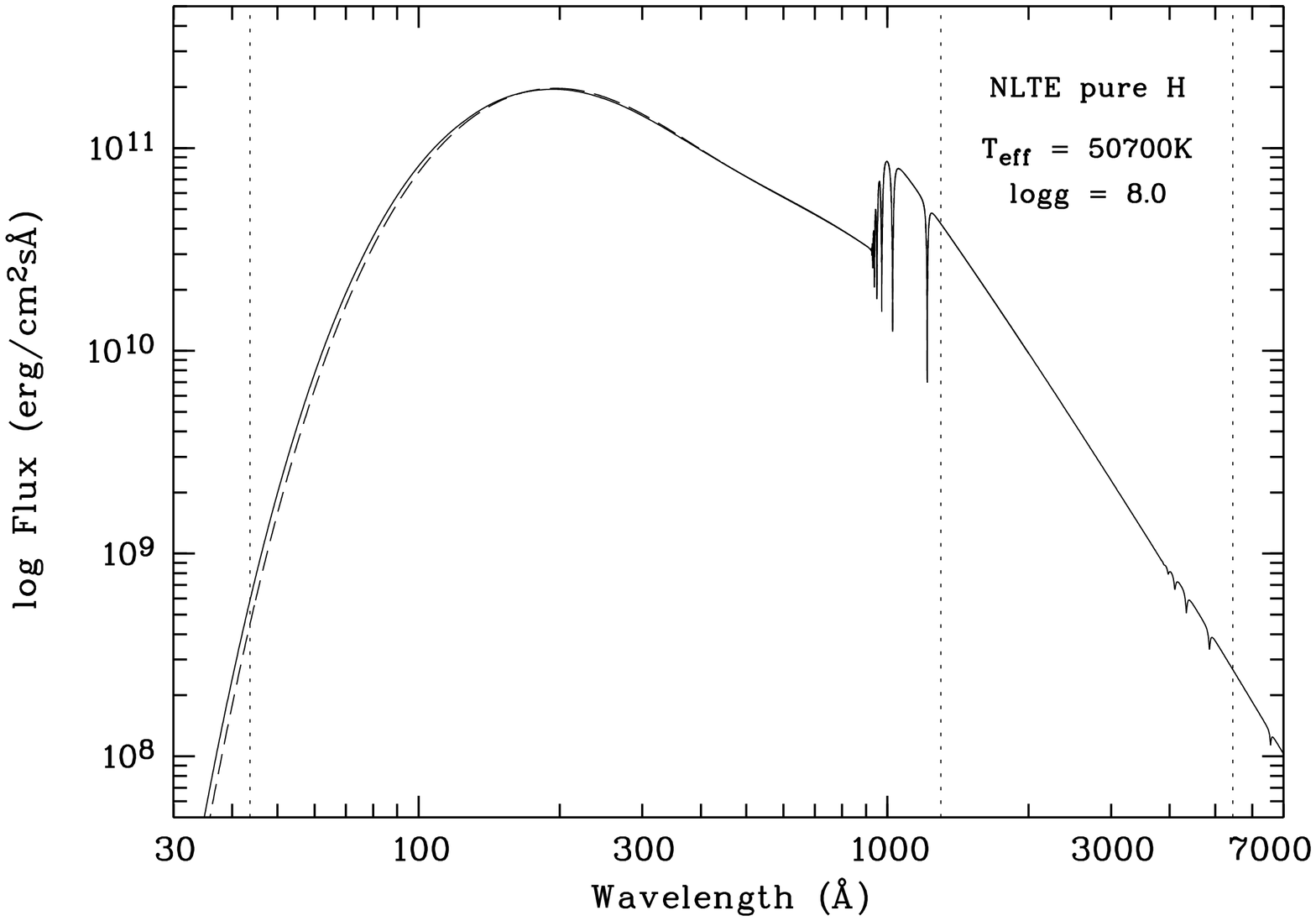}
\begin{minipage}[t]{5.0cm}
\vspace{-46.0mm}\hspace{24mm}
\includegraphics[width=3.5cm]{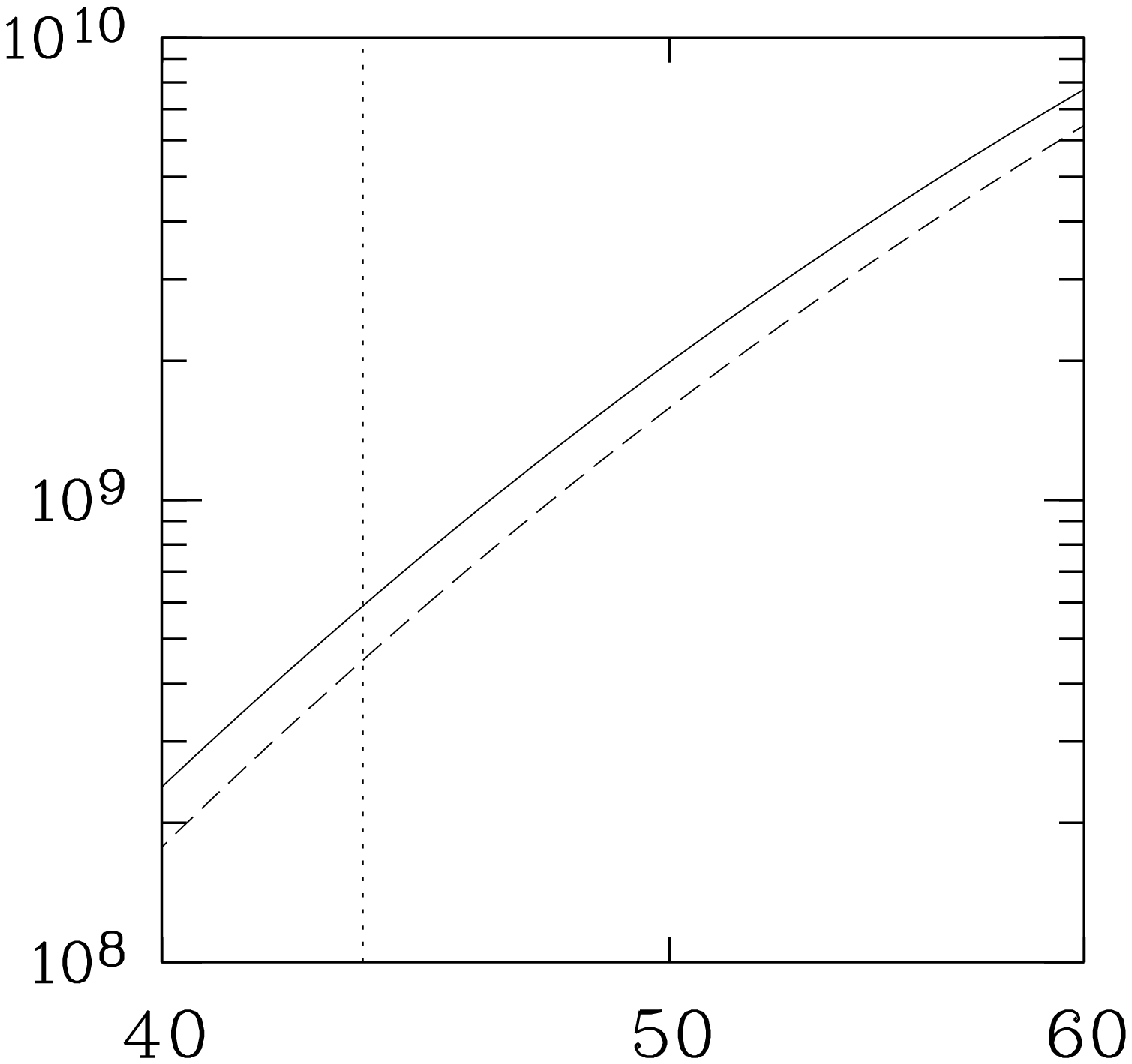}
\end{minipage}
\caption{NLTE pure hydrogen spectra for \teff=50700\,K and \logg=8.0,
calculated with TMAP using two versions of the bound-free and
free-free absorption coefficients, Seaton's approximation (dashed
curve) and the full Karszas \& Latter (1961) description (solid
curve). Both models yield practically identical results at
optical/ultraviolet wavelengths, but differ by a factor of 1.33 at
44\AA\ (vertical dotted line). The insert shows an expanded view. Two
further dotted lines mark $\lambda =1300$\AA\ and $\lambda =5450$\AA\
(see text). }
\label{fig:samplespec}
\end{figure}

\section{Model atmospheres and the case of HZ43}

In this Section, we describe the non-LTE model atmospheres used in our
analysis and summarize previous results on the photospheric parameters
of HZ43.

\subsection{NLTE models for pure hydrogen atmospheres}
\label{sec:nltespectra}

We employ \textit{TMAP}, the T\"ubingen NLTE Model Atmosphere Package
\citep{werneretal03,rauchdeetjen03}, in its most recent version for
the calculation of plane-parallel, static, pure-hydrogen models.
Hydrogen is represented by a model atom with the lowest levels up to
principal quantum number $n = 14$ treated in NLTE. The higher levels
are dissoluted rapidly towards the inner photosphere at the high $g$
considered here, e.g., to less than 9\% occupation probability at a
column density of only $10^{-5}\,\mathrm{g\,cm}^{-2}$ for $n=14$ and
the \teff~=~27900\,K and \logg~=~8.6 of Sirius\,B
\citep{hummermihalas88}. Hence, this number of NLTE levels is
sufficient. Two additional levels are treated in LTE. The NLTE levels
are fully coupled by radiative line transitions. An approximate
formula for the Stark broadening \citep{unsoeld68} is employed for the
calculation of their absorption coefficients. Collisional transitions
are considered between all levels. The atmospheres are calculated
within optical depths of $\log \tau = -8 \ldots +4$.  In order to
model the flux at the Lyman threshold precisely within
900\,--\,1000\,\AA, a narrow frequency grid was used with a spacing
equivalent to 0.1\,\AA.  For the synthetic spectra, we use the
line-broadening tables of \citet{lemke97}.

The short-wavelength continuum benefits from the use of $ff$-cross
sections from
\citet{sutherland98}\,\footnote{http://astro.uni-tuebingen.de/\raisebox{.2em}{\small
$\sim$}rauch/TMAP/UserGuide/ UserGuide.html} which replace the older
fit formula from \citet{mihalas67}. More importantly, the continuum
flux depends on the treatment of the $bf$-transitions. Using the full
formalism of \citet{karzaslatter61} raises the flux at 44\AA\ by 33\%
compared with spectra calculated with the older Seaton formula and
Gaunt factor (see Fig.~\ref{fig:samplespec}). We find very good
agreement of our results with the short-wavelength continuum of
\citet{barstowetal03}. Given the sensitivity of the soft X-ray fluxes
to details of the calculation, we caution against the indiscriminate
comparison of different versions of model spectra used over the
decades in the calibration of various satellite instruments sensitive
in the soft X-ray regime.

\begin{figure}[t]
\begin{center}
\includegraphics[width=8.6cm]{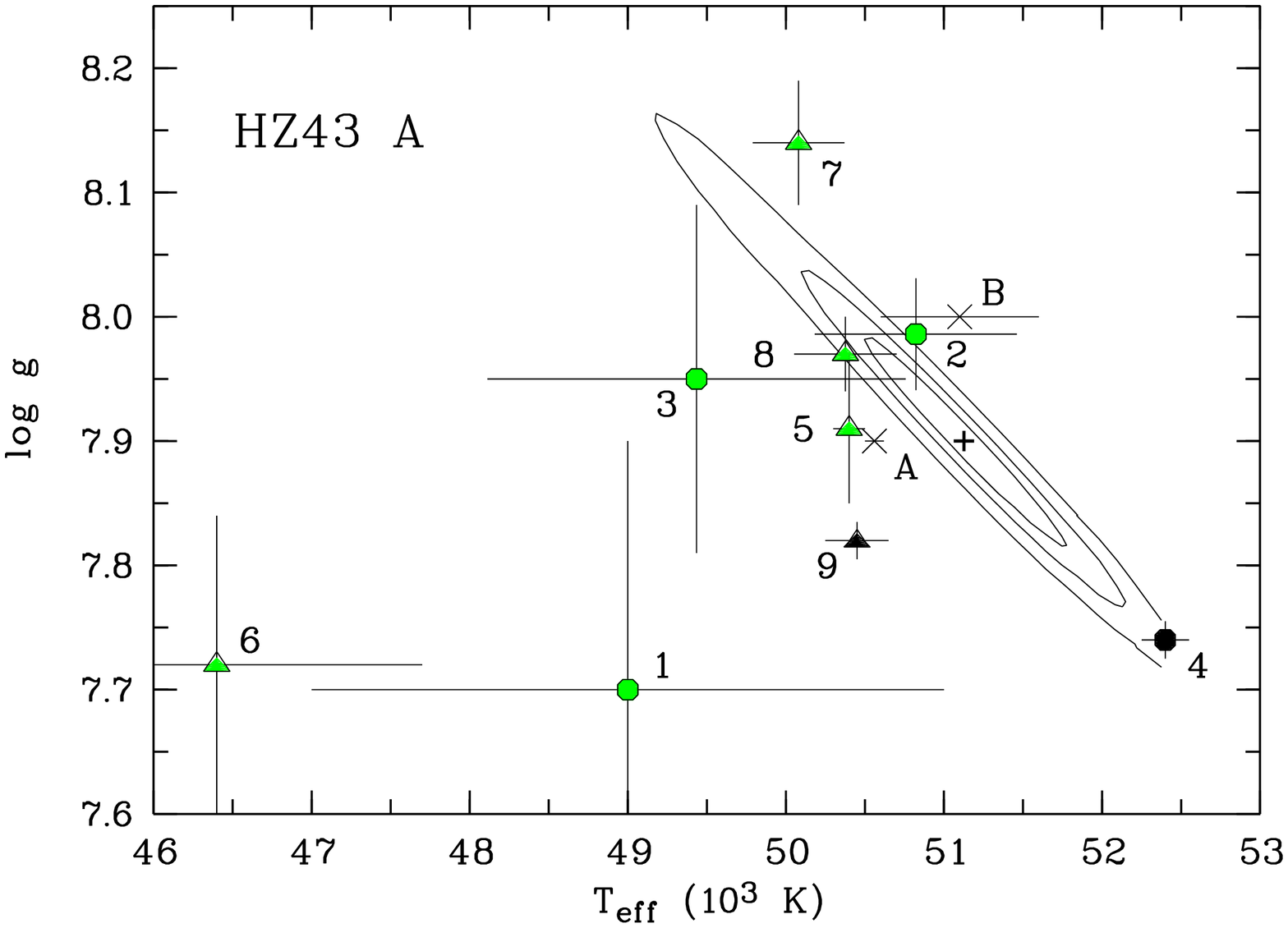}
\end{center}
\caption{Summary of \teff--\logg\ determinations for \ohz\ from fits
to the Balmer lines (circles), the Lyman lines (triangles), and the
\textit{EUVE} continuum (lying crosses):
1\,=\,\citet{napiwotzkietal93}; 2\,=\,\citet{finleyetal97};
3\,=\,\citet{barstowetal03}; 4\,=\,this work (STIS Balmer lines);
5\,=\,\citet{dupuisetal98} (\textit{ORFEUS});
6\,=\,\citet{barstowetal01} (\textit{HUT});
7\,=\,\citet{barstowetal01} (\textit{ORFEUS});
8\,=\,\citet{barstowetal03} (\textit{FUSE}); 9\,=\,this work
(\textit{FUSE}); A\,=\,\citet{barstowetal95} (\textit{EUVE}); and
B\,=\,\citet{vennesdupuis02} (\textit{EUVE}). The error ellipses
indicate the results of the present fits to the \chandra\ soft X-ray
continuum and refer to the 68\%, 90\% and 95\% confidence levels.}
\label{fig:hz43par}
\end{figure}

Another aspect suggested to be important at soft X-ray energies is the
suppression of the flux by Compton rather than Thomson
scattering. Although \citet{madej98}, claimed a sizeable effect,
\citet{suleimanovetal06} show that it is only at the 1\% level near
the carbon K-edge in \ohz\ and \osi. Since the Compton effect deforms
the model spectrum, we investigate the effect quantitatively below.

\subsection{The case of \ohz}

Figure~\ref{fig:hz43par} summarizes previous measurements of \teff\
and \logg\ of \ohz\ supplemented by our present results.  The
individual parameter combinations scatter over 6000\,K and 0.4~dex,
respectively, with no preference for any individual method to prefer a
particular region in parameter space. Systematic differences in the
deduced parameters from Lyman and Balmer lines were extensively
studied by Barstow and collaborators, who show that they are still
small near 50000\,K \citep{goodetal04}. The data points in
Fig.~\ref{fig:hz43par} cluster around \teff~=~50500\,K and
\logg~=~7.95, but it is not certain that such a poll bears physical
significance. The cause for the partly disparate data points is likely
systematic in nature and may have its origin in observational
problems, e.g., the presence of stray light in the line centers; in
differences of the fitting procedures, e.g., the way a quasi-continuum
is adjusted; in differences of models, e.g., between LTE and non-LTE
or between the individual versions of the line broadening theory; and
in physical differences as the unrecognized presence of trace metals
and their levitation
\citep{barstowetal03,bohlin00,chayeretal95,napiwotzki97,schuhetal02}. 
Clearly, the extremely good counting statistics in the spectra of
\ohz\ in all spectral ranges are not the only error source. The error
ellipses in Fig.~\ref{fig:hz43par} denote our result for \ohz\ based
on the interpretation of the \chandra\ soft X-ray spectrum. This
result takes systematic errors partly into account and is described
below.

\section{Results}

Model fits to the individual \chandra\ LETG+HRC spectra using the
nominal (November 2004) first order effective areas \anomi\ and the
nominal efficiencies of the higher orders relative to the first order
yield marginally acceptable reduced \chisqred~=~1.45 and 2.04 for
\rxj\ and \osi, respectively, but an enormous \chisqred~=~32.40 for
\ohz.  Following suggestions by others
\citep[e.g.][]{brajeromani02,drakeetal02,peaseetal00,peaseetal03}, we
interprete these deficiencies as caused by the remaining errors in
\anomi\ and search for a way to remove these defects. To this end, we
include both hot white dwarfs in a simultaneous fit together with the
neutron star \rxj, which allows us to tie \ohz\ to the flux level of
the latter and to the better defined effective areas at $\lambda <
40$\AA. In what follows, we discuss the normalizations of the model
spectra and the treatment of the interstellar absorption which is of
particular importance for \rxj\ and subsequently describe the combined
fit.

\subsection{Visual and EUV fluxes}

We require that our incident spectra comply with the findings in other
wavelength regions. For \ohz, we adopt the visual magnitude
$V=12.909$ \citep{bohlin00}, or equivalently adjust the model to the
HST/STIS spectral flux at 5450\AA, $f_\mathrm{5450}=
\ten{2.558}{-14}$\,\ergsa. In order to appropriately fix the
interstellar absorbing hydrogen column density, we adopt a mean flux
in the 475--495\AA\ region of $f_\mathrm{485} =
\ten{1.84}{-12}$\,\ergsa, which is the spectral flux of the
long-wavelength \textit{EUVE} spectrum used by \citet{wolffetal99} and
slightly exceeds that of \citet{dupuisetal95}. For \osi, we adopt
the 4600\AA\ HST/STIS flux $f_\mathrm{4600}=
\ten{2.860}{-12}$\,\ergsa\ taken from the carefully calibrated G430L
HST/STIS spectrum of \citet{barstowetal05} which puts us on his
optical flux scale. The mean 475--495\AA\ \textit{EUVE} flux of
\osi, $\sim \ten{6}{-15}$\,\ergsa, has a huge error, but a flux of
this order is consistent with the interstellar neutral hydrogen column
density reported by \citet{hebrardetal99}, \nh$~=\ten{(6.5\pm
2.0)}{17}$\,cm$^{-2}$. For \rxj, we use the visual flux determined by
\citet{kerkwijkkulkarni01}, $f_\mathrm{5000}=\ten{2.96}{-19}$\,\ergsa\
at 5000\AA, and account for extinction using $A_{V}=N_{\rm
HI}/(\ten{1.87}{21}x_\mathrm{H}$\,cm$^{-2}$) \citep{claytonetal03},
where \nh\ is the fitted interstellar column density of atomic
hydrogen in cm$^{-2}$ and \xh\ is the ionization fraction of hydrogen.

\subsection{Interstellar photoelectric absorption}

The strength of the interstellar carbon K-shell absorption edge in the
spectrum of \rxj\ is central to our discussion and requires that we
discuss this aspect to some detail. We note that the application of
the nominal LETG+HRC effective areas \anomi\ yields an unphysical
carbon jump of the \rxj\ spectrum in \emph{emission}, which 
corresponds to a negative carbon abundance \abc. The
appropriate treatment of carbon absorption allows us to fix the
correction to the LETG+HRC effective area on the long-wavelength side
of the carbon jump. This is the first step in our approach to an
internally consistent set of photon spectra for our three stars.

In treating interstellar photoelectric absorption, we largely follow
\citet{wilmsetal00}. We use the cross-sections of hydrogenic ions as
given by the analytic formula from Spitzer's textbook, the cross
sections of \citet{yanetal98} for neutral helium, and the cross
sections of \citet{balucinskaetal92} and \cite{verneryakovlev95} as
implemented in the most recent version of
XSPEC\footnote{http://heasarc.gsfc.nasa.gov/docs/xanadu/xspec} for the
heavier elements. For the short lines of sight to the objects
considered here, we neglect molecules.  We use the protosolar
abundances of \citet{lodders03} as representative of cosmic
abundances, except for carbon which is discussed below, and consider
the reduced effective cross-sections of atoms condensed in dust
grains.
The solar and cosmic abundance of carbon has long been considered to
be around 400\,ppm \citep[e.g.][]{holweger01}\footnote{We quote
abundances in parts per million (ppm) relative to hydrogen. C/H=245
(400)\,ppm corresponds to a logarithmic abundance of 8.39 (8.60).}, but
\citet{allendeprietoetal02} re-determined the photospheric solar
abundance to be 245\,ppm, while \citet{lodders03} suggested a
protosolar abundance of $288\pm 30$\,ppm which is based on Allende
Prieto's photospheric value and accounts for some elemental settling
in the Sun. Measurements of the gaseous carbon abundance in the
tenuous warm interstellar medium yield C/H $\simeq 140-161$\,ppm
\citep{cardellietal96,sofiaetal04}.  There is a dispute about how much
carbon is needed to explain the extinction properties of dust
grains. Synthesis of the extinction curve for translucent lines of
sight in terms of the properties of graphite, amorphous carbonaceous
grains, and silicate grains suggests that a minimum of 180\,ppm, and
more typically 250\,ppm, of carbon is needed to form grains with the
observed properties \citep{claytonetal03,zubkoetal04}, which suggests
a carbon abundance in the general interstellar medium somewhat larger
than protosolar. The earlier proposition of an interstellar carbon
abundance lower than solar photospheric \citep[e.g.][]{wilmsetal00}
has become unlikely with the reduction of the solar value. We
generously condense the available information into a range of the
total interstellar carbon abundance of $320\pm 80$\,ppm of which
$\sim60$\% is in dust grains. The size distribution of the grains is a
subject of dispute, too, since the ones deduced from the extinction
curves in different directions
\citep[e.g.][]{claytonetal03,zubkoetal04} differ from that of grains
of interstellar origin picked up in interplanetary space
\citep{frischetal99}. Larger grains produce less extinction and will
be less efficient X-ray absorbers. We give preference to the size
distribution derived from extinction curves and use a MRN distribution
\citep{mathisetal77,claytonetal03} with a cutoff at a maximum grain
radius of $0.5\,\mu$m as our best estimate to reproduce the effects of
carbonaceous grains. The influence of this parameter on our results is
\mbox{moderate}.

We fix the degrees of ionization of the interstellar hydrogen and
helium along the line of sight to our three stars. For \ohz, we use
\xh~=~0.10 and \xhe~=~0.40 \citep{wolffetal99}. For the other two stars,
we adopt \xh~=~0.25 and \xhe~=~0.40, as determined from pickup ions of
local interstellar origin and other studies
\citep{slavinfrisch02,gloecklergeiss04}. Doubly ionized helium is
disregarded following \citet{slavinfrisch02}.

\subsection{The LETG+HRC spectrum of \rxj}
\label{sec:rxj}

We followed \citet{burwitzetal03}, \citet{brajeromani02}, and
\citet{ponsetal02} in fitting the observed spectrum of \rxj\ from the
optical to the X-ray regime by the sum of two blackbodies with
temperatures k$T_\mathrm{spot}$ and k$T_\mathrm{star}$, of which the
hotter one (k$T_\mathrm{spot}$) describes the X-ray emitting spot. We
provisionally interprete the cooler component (k$T_\mathrm{star}$) as
the emission from the annulus representing the projected neutron star
surface outside the spot\,\footnote{Presently, the nature of the cool
component is still disputed and so is the angular stellar radius
derived from the blackbody assumption. A model in which an underlying
blackbody flux is reprocessed in a thin surface layer of hydrogen can
account for an enhancement of the optical flux relative to that at
shorter wavelengths, but only by a factor of at most two (S. Dreizler,
private communication) and, of course, it can not exceed the blackbody
flux at its own temperature. A way to dissociate the optical component
from the neutron star as a light source would be the presence of
optical light emitting circumstellar matter.}.

\begin{figure}[t]
\includegraphics[width=8.8cm]{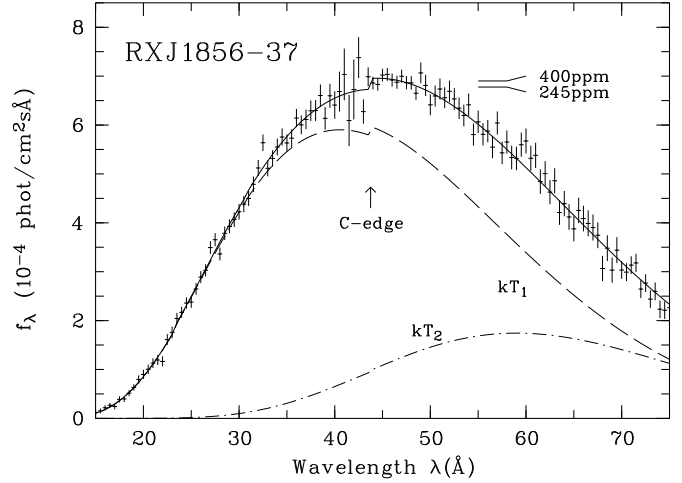}
\caption{\chandra\ LETG+HRC spectrum of \rxj\ binned to 0.5\AA,
derived with the effective area at wavelengths longwards of the carbon
edge corrected to fit an abundance of carbon in the interstellar
medium of $a(C)=245-400$\,ppm with 60\% in dust grains. The
two-temperature fit shown is for 320\,ppm, with the upper level of the
step indicated for 245 and 400\,ppm, too.  The fit has a \chisq$=97.7$
for 115 dof (see Sect. 4.5). The model parameters are quoted in
Table~\ref{tab:results}.}
\label{fig:rxj}
\end{figure}

We fit the summed model spectrum for the first to the sixth
diffraction orders to the \chandra\ LETG+HRC-S count rate spectra of
\rxj\ in the positive and negative dispersion directions.
Fig.~\ref{fig:rxj} shows the corresponding summed photon spectrum,
which allows to judge the strength of the interstellar carbon
absorption edge more clearly than the count rate spectrum with its
superposed effective area structure. The photon spectrum is created
from the observed spectrum by subtracting the \mbox{$2^\mathrm{nd}$ to
$6^\mathrm{th}$} order model count rates from the data (in \ctsa) and
dividing by the first-order effective area. The expected strength of
the absorption edge is obtained if the effective areas shortwards of
43.7\AA\ are accepted and longwards are moderately reduced. We perform
fits to the wavelength region 15--39\AA\ and anchor the fit at long
wavelengths where the higher orders dominate. This yields a data/model
ratio different from unity in the 44.0--48.5\AA\ range which we
interprete as the mean factor $\alpha_1$ by which the effective area
in this interval has to be adjusted in order to obtain a perfect fit
with these wavelengths included. From best fits with \abc\ between 245
and 400\,ppm and 60\% of the carbon in dust grains, we obtain
$\alpha_1\equiv \langle A\rangle/\langle A^0\rangle =0.875\pm 0.012$,
where the angular brackets refer to averages over the 44--48.5\AA\
interval. An additional and larger uncertainty arises from the errors
in the fit parameters, notably k$T_\mathrm{star}$, determined for an
increase in \chisq\ by +1 over its minimum $\chi^2_\mathrm{min}=38.7$
for 44 dof in the 15--39\AA\ range. Merging the two results yields
$\alpha_1 =0.871\pm 0.038$. This factor applies within its error to
both the negative and the positive dispersion directions.
Figure~\ref{fig:rxj} shows the derived summed spectrum of both sides
using effective areas $A=\alpha(\lambda)\,$\anomi\ where the
wavelength-dependent adjustment factor $\alpha$ is derived below and
is forced to match $\alpha_1$ between 44 to 48.5\AA. Our photon
spectrum is very similar to that presented by \citet[][see their
Fig.~1]{drakeetal02}, who evidently used similarly adjusted effective
areas. Our final fit with the adjusted areas extends over the spectral
range 15--74\,\AA\ and has \chisq$~=~97.7$ for 115 dof. This spectral
range avoids the shortest wavelengths with low flux and the longest
wavelengths where the higher orders become important and finally
dominate.

\begin{figure*}[t]
\centerline{
\includegraphics[width=17.3cm]{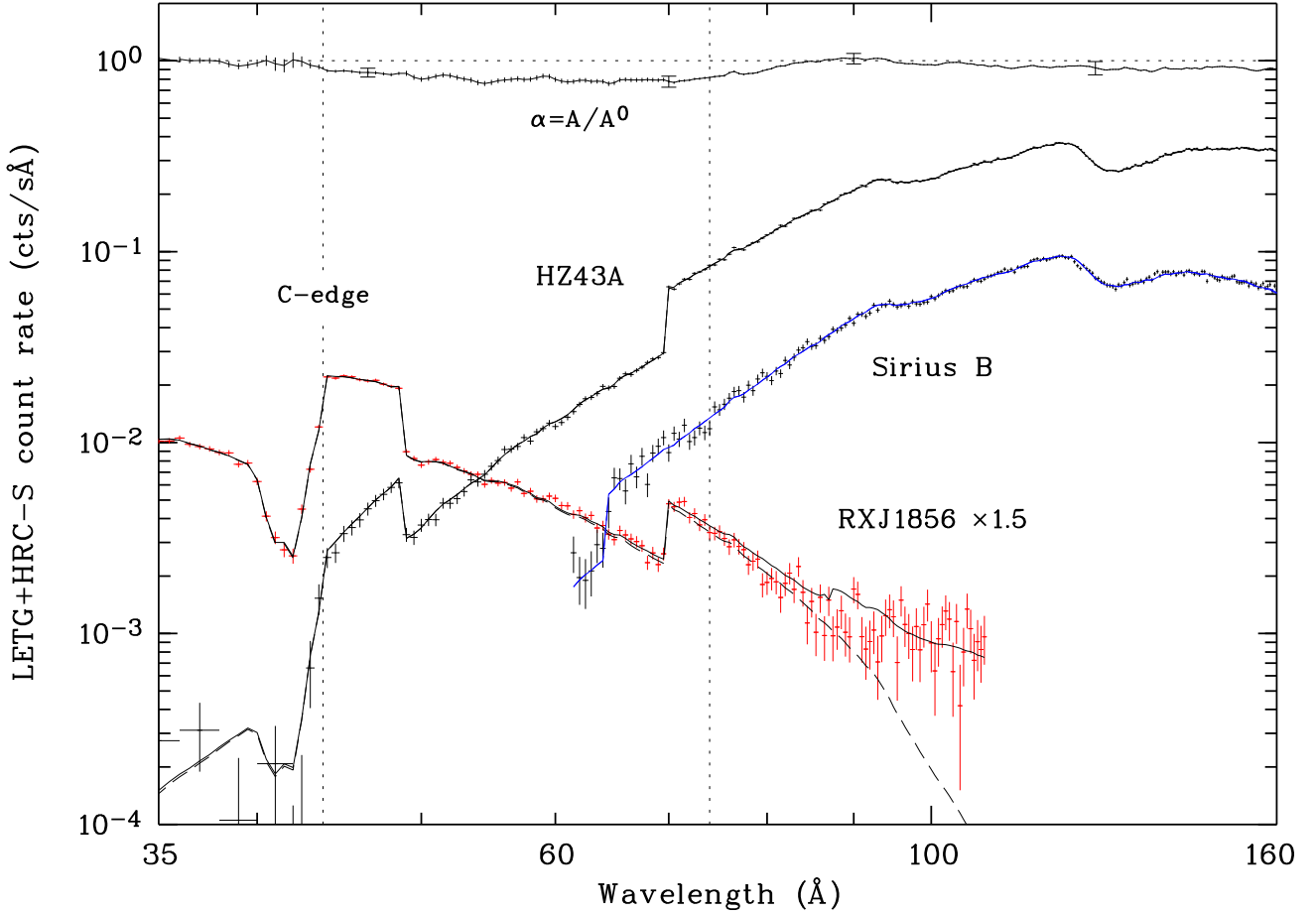}}
\caption{Simultaneous fit of \rxj, \ohz, and \osi\ in the wavelength
ranges marked by vertical dotted lines (see Sect. 4.4.2). The LETG
spectra binned to 0.5\AA\ are shown as data points, the corresponding
best-fit models as solid curves, and the first-order contributions as
dashed curves. The area correction function $\alpha$ is shown at the
top. It converts the nominal LETG+HRC-S first-order effective area
$A^0$ of the November 2004 release into the adjusted area $A$ used in
this paper. Systematic uncertainties in $\alpha$ are indicated by
error bars at 46, 70, 90, and 125\AA. The steps in the count rate
spectra of \ohz\ and \rxj\ at 49 and 69\AA\ result from the detector
gaps. \osi\ was observed off axis and its gaps are located differently
(see text).}
\label{fig:combined}
\end{figure*}

\subsection{Simultaneous fit to the spectra of several stars}

\subsubsection{General approach}

The method for the determination of the area correction function
$\alpha(\lambda)$ is based on the simultaneous fit to $n\ge 2$ stars
with $N$ data points each, centered at wavelengths $\lambda_1$ to
$\lambda_\mathrm{N}$. Because of a lack of suitable calibrators, we
apply the method for $n=2$ only. The standard \chisq\ which expresses
the differences between the data $y_\mathrm{ij}$ with statistical
errors $\sigma_\mathrm{ij}$ and the model contributions
$m_\mathrm{ijk}$ in orders k=1 to 6 (all in \ctsa) is
\begin{equation}
\chi^2 = \sum_\mathrm{j=1}^\mathrm{n}\sum_\mathrm{i=1}^\mathrm{N}
\frac{(y_\mathrm{ij}-\sum_\mathrm{k=1}^6m_\mathrm{ijk})^2}{\sigma_\mathrm{ij}^2}
\label{eq:chi1}
\end{equation}
with orders higher than k=6 entirely negligible for the objects considered.
As usual, the model contributions are calculated as
\begin{equation}
m_\mathrm{ijk}=\frac{1}{\mathrm{k}}f_\mathrm{j}
\left(\frac{\lambda_\mathrm{i}}{\mathrm{k}}\right)
\alpha\left(\frac{\lambda_\mathrm{i}}{\mathrm{k}}\right)
A^0\left(\frac{\lambda_\mathrm{i}}{\mathrm{k}}\right)
\varepsilon_\mathrm{k}\left(\frac{\lambda_\mathrm{i}}{\mathrm{k}}\right)=
\alpha\left(\frac{\lambda_\mathrm{i}}{\mathrm{k}}\right)m^0_\mathrm{ijk}
\label{eq:chi2}
\end{equation}
where $f_\mathrm{j}(\lambda_\mathrm{i})$ is the incident spectral flux
of star j in \photsa, $\varepsilon_\mathrm{k}(\lambda_\mathrm{i})$ the
efficiency of the k-th order relative to the first order, and
$m^0_\mathrm{ijk}$ are the model contributions calculated with the
nominal effective areas. We correct the observed count rates for the
respective model contributions in the second to sixth order to yield
$y'_\mathrm{ij}=y_\mathrm{ij}-\sum_\mathrm{k=2}^6
\alpha(\lambda_\mathrm{i}/\mathrm{k})m^0_\mathrm{ijk}$. Since
$\alpha(\lambda_\mathrm{i}/\mathrm{k})$ is involved, the
transformation requires, in principle, an iterative approach.  While
this is easily implemented in the iterative procedure to minimize
\chisq, the influence of this procedure on the derivation of the
$\alpha_\mathrm{i}=\alpha(\lambda_\mathrm{i})$ is minute because
\mbox{(i) the} adjusted areas affect the second and third orders
only at $\lambda>88$\AA\ and $\lambda>132$\AA, respectively,
\mbox{(ii) \rxj} is not affected because we use its data only at
$\lambda<74$\AA\ (see below), and \mbox{(iii) the} contributions of
the higher orders to the spectra of \ohz\ and \osi\ are close to
negligible. Hence the determination of the $\alpha_\mathrm{i}$ is
straightforward. We substitute the $y_\mathrm{ij}$ in
Eq.~(\ref{eq:chi1}) by the $y'_\mathrm{ij}$\,, form the first-order
count rate ratio of data vs. nominal model
$r_\mathrm{ij}=y'_\mathrm{ij}/m^0_\mathrm{ij1}$ with errors
$\rho_\mathrm{ij}=\sigma_\mathrm{ij}/m^0_\mathrm{ij1}$, and find that
the $\alpha_\mathrm{i}$ are determined by minimizing the differences
between the individual ratios $r_\mathrm{ij}$\,,
\begin{equation}
\chi^2 = \sum_\mathrm{j=1}^\mathrm{n}\sum_\mathrm{i=1}^\mathrm{N}
\frac{(y'_\mathrm{ij}-\alpha_\mathrm{i}m^0_\mathrm{ij1})^2}{\sigma_\mathrm{i}^2}
= \sum_\mathrm{j=1}^\mathrm{n}\sum_\mathrm{i=1}^\mathrm{N}
\frac{(r_\mathrm{ij}-\alpha_\mathrm{i})^2}{\rho_\mathrm{ij}^2}\,.
\label{eq:chi3}
\end{equation}
For $n=2$ stars, we compute $\alpha_\mathrm{i}$ as the weighted mean
of $r_\mathrm{i1}$ and $r_\mathrm{i2}$,
\begin{equation}
\alpha_\mathrm{i} = \frac{\rho_\mathrm{i2}^2 r_\mathrm{i1} + 
\rho_\mathrm{i1}^2 r_\mathrm{i2}}{\rho_\mathrm{i1}^2+\rho_\mathrm{i2}^2}\,,
\label{eq:chi4}
\end{equation}
and obtain the final form of the \chisq\ to be minimized as
\begin{equation}
\chi^2 = \sum_\mathrm{i=1}^\mathrm{N}
\frac{(r_\mathrm{i1}-r_\mathrm{i2})^2}{\rho_\mathrm{i1}^2+\rho_\mathrm{i2}^2}.
\label{eq:chi5}
\end{equation}
The expectation value of \chisq\ of Eq.~(\ref{eq:chi5}) is $N-p$,
since with $2N$ data points we determine $p$ parameters in the models
$m_1$ and $m_2$ and $N$ values of $\alpha_\mathrm{i}$.
The approach becomes insensitive to the $\alpha(\lambda_\mathrm{i})$
if the contributions of the higher orders at $\lambda_\mathrm{i}$
become large, because the relative errors of the $y'_\mathrm{ij}$
increase correspondingly. Minimizing \chisq\ with the
$\alpha_\mathrm{i}$ included may only be marginally stable if the
parameters allow a bulging deformation of both model spectra and a
runaway of the $\alpha_\mathrm{i}$. The white dwarf models are
resistant against such deformation, however, and
$\alpha(\lambda_\mathrm{i})$ is, furthermore, fixed to $\alpha_1 =
0.87$ in the interval 44--48.5\AA. In our application of the method,
the fitted spectra of \ohz\ and \rxj\ are tied down in the optical
\textit{and} at $\sim46$\AA, while the spectrum of \osi\ is tied to
\ohz.

We have opted to fit the sum of the model contributions for the
positive and negative dispersion directions to the sum of the
respective observed spectra. The so-defined model spectrum allows us
to derive the combined correction function $\alpha$ from the combined
observed spectrum as well as the individual $\alpha_\mathrm{neg}$ and
$\alpha_\mathrm{pos}$ from the observed spectra for the negative and
positive dispersion directions, respectively. The combined $\alpha$ is
the area-weighted mean of $\alpha_\mathrm{neg}$ and
$\alpha_\mathrm{pos}$,
\begin{equation}
\alpha A^0 = \alpha_\mathrm{neg} A^0_\mathrm{neg}+\alpha_\mathrm{pos}
A^0_\mathrm{pos}\, ,
\end{equation}
with $A^0=A^0_\mathrm{neg}+A^0_\mathrm{pos}$.  Multiplication of the
nominal area with the respective $\alpha$ yields the adjusted areas
$A_\mathrm{neg}=\alpha_\mathrm{neg} A^0_\mathrm{neg}$,
$A_\mathrm{pos}=\alpha_\mathrm{pos} A^0_\mathrm{pos}$, and $A =\alpha
A^0$.

\begin{table}[t]
\caption{Parameters of \ohz, \osi, and \rxj\ based on the simultaneous
fit of our model spectra to the LETG+HRC count rate spectra in the
wavelength intervals given. The quoted 1-$\sigma$ ($\Delta \chi^2=+1$)
errors are correlated and derived from fits with the other parameters
for each object kept free. The letter $f$ indicates: fixed.}
\label{tab:results}
\begin{tabular}{@{\hspace{5mm}}l@{\hspace{15mm}}cc} 
\hline \hline \noalign{\smallskip}

Parameter & Value$\pm $Error \\ 
             
\noalign{\smallskip} \hline
\noalign{\smallskip}
\multicolumn{2}{l}{\textit{\hspace{-1.5mm}(a) \ohz\ ($\lambda=45-160$\,\AA)}} &\\[0.2ex]
\teff\ (K)             &\hspace{-3mm}$51126\pm 660$	        & \\
\logg\                 & $7.90\pm 0.08$         & \\
$R^2/d^2~~(10^{-23}$)  &\hspace{-0mm}$3.011\pm 0.038$                 &  \\
\nh\ ($10^{17}$\,cm$^{-2}$) &\hspace{-1mm}  $8.91\pm 0.37$\\[0.8ex]
\multicolumn{2}{l}{\textit{\hspace{-1.5mm}(b) \osi\ ($\lambda=74-160$\,\AA)}} & \\[0.2ex]
\teff\ (K)             & \hspace{-3mm}$24923\pm 115$	        & \\
\logg\                 & \hspace{-3mm}$8.6~~f~^1$         & \\
$R^2/d^2~~(10^{-21})$  & \hspace{-1mm} $4.877\pm 0.036$                &  \\
\nh\ ($10^{17}$\,cm$^{-2}$) & \hspace{1.0mm}  $6.5\pm 2.0~^2$\\[0.8ex]
\multicolumn{2}{l}{\textit{\hspace{-1.5mm}(c) \rxj\ ($\lambda=15-74$\,\AA)}} & \\[0.2ex]
k$T_\mathrm{spot}$ (eV)            & \hspace{-1.6mm}$62.83\pm 0.41$	        & \\
k$T_\mathrm{star}$ (eV)            & \hspace{-1.9mm}$32.26\pm 0.72$	        & \\
$R_1/d$ (km/pc)        & \hspace{-0.5mm}$0.0378\pm 0.0003$        & \\
$R_2/d$ (km/pc)        & \hspace{-0.5mm}$0.1371\pm 0.0010$         & \\
\nh\ ($10^{20}$\,cm$^{-2}$) & \hspace{-1mm}  $1.10\pm 0.03$\\[0.2ex]
\noalign{\smallskip} \hline  \noalign{\smallskip}      
\end{tabular}

$^1$ Based on \citet{barstowetal05, holbergetal98}\\ $^2$
\citet{hebrardetal99}. Our fit is required to stay within the
1-$\sigma$ error.
\end{table}

\subsubsection{Combined fit to \rxj, \ohz, and \osi}

The number of stars with sufficiently well-known incident spectra is
extremely restricted. We fit, therefore, \rxj\ alone for $15 \le
\lambda < 39$\AA, \rxj\ and \ohz\ simultaneously for $44 \le \lambda
< 74$\AA, and \osi\ and \ohz\ simultaneously for $74\le \lambda
\le 160$\AA.  We calculated a grand~~\chisq\ for the combined fit of
the three models to the three observed spectra with 518 data points
including three points in the visual and the two 485\AA\ fluxes of
\ohz\ and \osi\ (the latter chosen to reproduce the
\citet{hebrardetal99} value of \nh\ within its error). The number of
free parameters includes the 12 listed in Table~\ref{tab:results} plus
the 232 \mbox{values $\alpha_\mathrm{i}$}.  Hence, there are 274 dof.
Figure~\ref{fig:combined} displays the results of the simultaneous fit
to the three stars and Table~\ref{tab:results} gives the best-fit
parameters. The quoted \mbox{1-$\sigma$} errors are derived from
variation of the respective parameter with all other parameters kept
free and refer \mbox{to $\Delta\chi^2=+1$}. The only parameter kept
fixed in the fit was the gravity of \osi, \logg$~=8.6$, which is
reasonably well determined from the angular radius, the
\textit{Hipparcos} distance and the gravitational redshift
\citep{barstowetal05, holbergetal98}. The grand \chisq\ for the
simultaneous fit is \mbox{282.9 for 274 dof} (\,\chisqred~=~1.03)
which is close to perfect.

The self-consistently determined multiplicative correction function
$\alpha(\lambda_\mathrm{i})=A/A^0$ for the sum of negative and
positive dispersion orders is shown at the top of
Fig.~\ref{fig:combined}. It converts the nominal LETG+HRC-S effective
area $A^0$ of the November 2004 release to the area $A$ used in this
paper and is valid for the best-fit \logg=7.90 of \ohz. It has been
smoothed with a box car over three wavelength bins (i.e., a total
width of 1.5\AA) in order to reduce the influence of statistical
fluctuations in the individual spectra. The correction is consistent
with unity at $\lambda<44$\AA, but reaches a low of about 0.77 at
65\AA, consistent with Pease's et al. (2003) expectation that the
error in the 60--80\AA\ range might reach 25\%. At longer wavelengths,
$\alpha(\lambda_\mathrm{i})$ moves closer to unity again. For \logg\
values of \ohz\ other than 7.90, the whole curve moves up and down a
bit with a pivot point at $\alpha(46\AA)=0.87\pm0.04$. In order to
mark the extent of this variation, we have included the systematic
errors at 46, 70, 90, and 125\AA\ with an upper (lower) cross bar
signifying the location of $\alpha(\lambda_\mathrm{i})$ for
$\alpha_1=0.91 (0.83)$ and \logg=8.00 (7.80) for \ohz, respectively.
None of the other parameters has a similarly large influence. In
summary, we conclude that the necessary adjustment to the LETG+HRC-S
effective area relative to that at $\lambda\la 40$\AA\ is pronounced
in the 50--80\AA\ region, but small longwards of 85\AA. Its systematic
uncertainty is $\pm 5$\% at $\lambda \ga 80$\AA\ from the uncertainty
in \logg\ of \ohz\ and another 5\% at all wavelengths $\lambda \ga
44$\AA\ from the error in $\alpha_1$. The fact that \logg\ of \osi\
is better defined than that of \ohz\ does not help because the
spectrum of \osi\ does not extend to 44\AA\ where it could be tied
down.

Our derived areas for the negative and positive dispersion directions,
$A_\mathrm{neg}= \alpha_\mathrm{neg} A^0_\mathrm{neg}$ and
$A_\mathrm{pos}= \alpha_\mathrm{pos} A^0_\mathrm{pos}$, are close to
the SRON areas suggested previously by J. Kaastra \citep[][and
J. Kaastra, private communication]{peaseetal00} between 44 and 80\AA\
and close to the nominal CXC areas at longer wavelengths. Our results
are provided in Tables~A1 to A3\,\footnote{Tables A1 to A3 are
available only electronically at
http://cdsweb.u-strasbg.fr/cgi-bin/qcat?J/A+A/.} for the preferred
\logg\ of \ohz\ of 7.90 as well as for 7.80 and 8.00, indicating the
systematic errors of our fit. The following systematic uncertainties
are not included in these Tables: (i) the normalization uncertainty at
$\sim46$\AA\ as expressed by $\alpha_1 = 0.87\pm 0.04$; (ii) the
possible degradation in the HRC response with time; and (iii) the
absolute error in the nominal areas $A^0$ at short wavelengths.

As noted above, our empirically adjusted effective areas are strictly
valid only for the epoch JD\,2452130. The possible variation of the
HRC sensitivity since the begin of the mission is small, however,
0.015 in $\alpha(\lambda_\mathrm{i})$ at long wavelengths and possibly
as much as 0.050 shortwards of about 50\AA\
(Fig.~\ref{fig:timevar}). Nevertheless, the possibility of a slow
degradation in $A_\mathrm{neg}$ and $A_\mathrm{pos}$ should be kept in
mind: it ultimately limits the accuracy of parameters determined
from fits to the long wavelength sections of LETG+HRC spectra.

\begin{figure}[t]
\includegraphics[width=8.7cm]{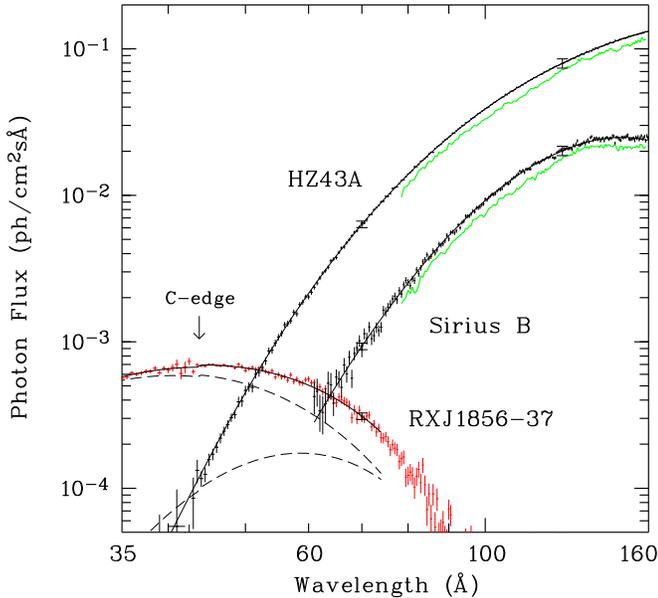}
\caption{Photon spectra of \rxj, \ohz, and \osi\ based on our analysis
of the LETG+HRC-S observations (data points) and best fit model
spectra with parameters as given in Table~\ref{tab:results} (solid
curves). The spectrum of \rxj\ is the same as in Fig.~\ref{fig:rxj}
with the two blackbody components shown separately as dashed
curves. Longwards of 78\AA, we have included the \textit{EUVE} spectra
of the two white dwarfs (green/gray curves).  }
\label{fig:phot}
\end{figure}

\begin{table}[t]
\caption{Soft X-ray fluxes as defined by the best-fit absorbed photon
spectra in Fig.~\ref{fig:phot}. The numbers in brackets are the
systematic errors of the LETG+HRC-S area correction in percent (see
text).}
\label{tab:fluxes}
\begin{tabular}{rccc} 
\hline \hline \noalign{\smallskip}

$\lambda~~$ & \ohz\           &  \osi\                & \rxj\ \\
 (\AA)    & \multicolumn{3}{c}{\hspace{3mm}photons~cm$^{-2}$s$^{-1}$\AA$^{-1}$}       \\
             
\noalign{\smallskip} \hline
\noalign{\smallskip}
 20 &                       &                       & $\ten{8.54}{-5}~~(5)$ \\
 30 &                       &                       & $\ten{4.32}{-4}~~(5)$ \\
 40 &                       &                       & $\ten{6.54}{-4}~~(5)$ \\
 43 &                       &                       & $\ten{6.70}{-4}~~(5)$ \\
 44 &                       &                       & $\ten{6.94}{-4}~~(5)$ \\
 48 & \hspace*{3mm}$\ten{3.01}{-4}~~(5)$~$^{1)}$ &  & $\ten{6.81}{-4}~~(5)$\\
 60 & $\ten{2.14}{-3}~~(5)$ & $\ten{2.49}{-4}~~(6)$ & $\ten{5.06}{-4}~~(7)$ \\
 70 & $\ten{6.33}{-3}~~(5)$ & $\ten{9.57}{-4}~~(6)$ & $\ten{3.11}{-4}~~(7)$ \\
 80 & $\ten{1.39}{-2}~~(6)$ & $\ten{2.58}{-3}~~(6)$ & \\
 90 & $\ten{2.49}{-2}~~(6)$ & $\ten{5.43}{-3}~~(6)$ & \\
100 & $\ten{3.89}{-2}~~(6)$ & $\ten{9.47}{-3}~~(6)$ & \\
125 & $\ten{7.96}{-2}~~(7)$ & $\ten{2.10}{-2}~~(7)$ & \\
160 & $\ten{1.31}{-1}~~(7)$ & $\ten{2.55}{-2}~~(7)$ & \\
\noalign{\smallskip} \hline 
\noalign{\smallskip}
    & \multicolumn{3}{c}{\hspace{3mm}erg~cm$^{-2}$s$^{-1}$\AA$^{-1}$}  \\     
\noalign{\smallskip} \hline      
\noalign{\smallskip}
1300 & $\ten{4.052}{-12}$ & $\ten{1.258}{-10}$&$\ten{5.07}{-17}~^{2)}$\\
4600 & & $\ten{2.860}{-12}$ &  \\
5000 & & & $\ten{2.96}{-19}~^{3)}$\\
5450 & $\ten{2.558}{-14}$ &  & \\
\noalign{\smallskip} \hline  
\end{tabular}
\noindent $^{1)}$ The Compton effect may depress this flux by
$\sim1$\%.\\ $^{2)}$  The unabsorbed flux is
$\ten{6.27}{-17}$\,\ergsa\\ $^{3)}$ The unabsorbed flux is
$\ten{3.18}{-19}$\,\ergsa
\end{table}

\subsubsection{The photon spectra of \ohz, \osi, and \rxj}

The incident photon spectra of the three stars fitted simultaneously
are obtained by dividing the count rate spectra of
Fig.~\ref{fig:combined} by our effective areas. Their errors are
dominated by the systematic errors as discussed in the last Section.
Fig.~\ref{fig:phot} shows the resulting photon spectra with the
systematic uncertainties indicated again by additional error bars at
60 and 125\AA. The spectral fluxes at selected wavelengths are listed
in Table~\ref{tab:fluxes} together with their systematic errors
arising from $\alpha$ in percent given in brackets. The complete
spectra are provided in Tables A4 to A6\,\footnote{Tables A4 to A6 are
available only electronically at
http://cdsweb.u-strasbg.fr/cgi-bin/qcat?J/A+A/.}. The systematic
errors listed near the end of the last Section under (ii) and (iii)
are not included. The bottom three lines of Table~\ref{tab:fluxes}
contains the optical fluxes to which the spectra were normalized. The
fourth line from the bottom gives the predicted fluxes at 1300\AA. In
the case of \ohz, the quoted flux agrees to better than 0.3\% with
that of the HST/STIS spectrum. For \osi, the 1300\AA\ model flux
agrees with that predicted by \citet{vennesdupuis02} and both exceed
the IUE flux which is known to be too low \citep{massafitzpatrick00}.

\subsection{Results for individual stars}

We now discuss the results for the individual stars taking the
systematic errors as far as possible into account. These fits are
characterized by \chisq=97.7 for \rxj\ in the interval
\mbox{15--74\AA} with 115 dof, \chisq=183.9 for \ohz\ in
45.5--160\AA\ with 229 dof, and \chisq=209.4 for \osi\ in
61.5--160\AA\ with 197 dof. All fits are substantially improved over
those without the area correction and can all be considered 'good' in
a statistical sense. This result is not subject to the binning chosen
as in previous fits to the LETG+HRC spectrum of \rxj\
\citep[e.g.][]{brajeromani02}. The combined \chisq\ for \ohz\ and
\osi\ is 393.3 for 426 dof and the slight disparity between the two
stars results form the better statistics of \ohz\ which drives the
fit. 

\subsubsection{\ohz}

The errors of the individual parameters listed in
Table~\ref{tab:results} are, in part, highly correlated. This is
particularly true of \teff\ and \logg\ for \ohz. The corresponding
error ellipses for the 68\%, 90\% and 95\% confidence limits are
overplotted in Fig.~\ref{fig:hz43par}. Gratifyingly, the best-fit
values are in good agreement with results obtained from the optical
and ultraviolet regimes. The large extent along the major axis of the
ellipses reflects the correlation between the results for \teff\ and
\logg. The soft X-ray flux stays essentially constant along the major
axis and the narrow width in the perpendicular direction signifies the
rapid deterioration of the fit if the model does not meet the observed
flux. Hence, the soft X-ray flux in Table~\ref{tab:fluxes} is better
defined by the fit than either \teff\ or \logg. Parameter combinations
like those in the lower left of Fig.~\ref{fig:hz43par} can safely be
excluded since they yield much too low a flux, parameter combinations
in the upper right would yield too large a flux.  This is an important
result for the future calibration of instruments sensitive in this
spectral regime: it is more appropriate to quote soft X-ray fluxes
than an effective temperature.

\begin{figure}[t]
\vspace*{0.4mm}
\includegraphics[width=8.8cm]{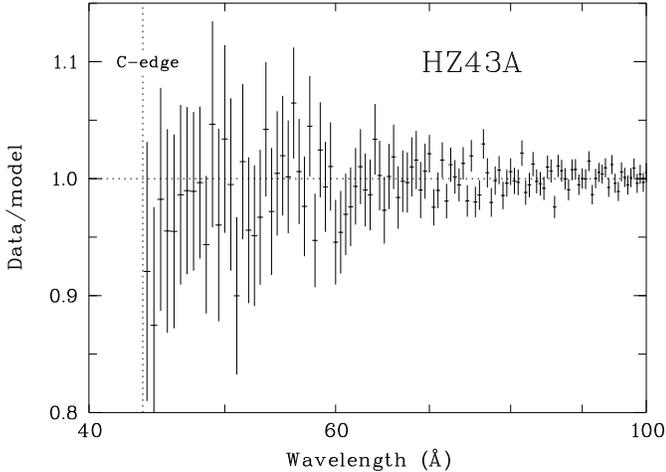}
\caption{Data vs. model flux ratio of the spectrum of \ohz\ for
$\lambda>43.5$\AA. The maximum depression of the spectrum by the
Compton effect near the carbon K-edge at 43.7\AA\ is at the percent
level.}
\label{fig:compton}
\end{figure}

The angular radius $R/d$ of \ohz\ fixes the distance $d$ for a given
radius $R$. Adopting \citet{wood95} stellar models with a thick
hydrogen envelope provides an independent relation and yields
$M=0.64\pm0.05\,M_\odot$, $R=\ten{(1.04\pm0.08)}{9}$\,cm, and
$d=61.3\pm5.5$\,pc with errors which account for the systematic
uncertainty due to the full gravity range \logg$~=7.90\pm0.10$. The
gravitational redshift predicted for these parameters is
$\upsilon_\mathrm{g}=28\pm 4$\,km\,s$^{-1}$. The observed
$\upsilon_\mathrm{g}=30.1\pm1.5$\,km\,s$^{-1}$ \citep{reid96} suggests
the slightly more restricted range $M=0.67\pm0.02\,M_\odot$
\citep{dupuisetal98}, with $R=\ten{(9.92\pm0.34)}{8}$\,cm,
log\,$g=7.95\pm 0.04$, and $d=59\pm2$\,pc (1-$\sigma$ errors). Both
distance values are consistent with the trigonometric parallax
\mbox{$\pi_\mathrm{abs}=15.3\pm2.9$}\,mas ($d=65\pm ^{15}_{10}$\,pc) from
the Yale Catalog of Trigonometric Parallaxes \citep{vanaltenaetal01}.

In principle, the LETG+HRC data of \ohz\ can help to decide whether
the short-wavelength spectra of hot white dwarfs are steepened by the
fact that electron scattering is no longer in the Thomson limit and
must be correctly described by the Compton effect which degrades the
energies of the scattered photons. In Fig.~\ref{fig:compton}, we show
the data/model ratio of the photon spectrum of \ohz\ from
Fig.~\ref{fig:phot} for $\lambda \ge 44$\AA. The data points are all
more or less consistent with a ratio equal to unity: $1.0005\pm
0.0014$ for $\lambda = 65-100$\AA; $0.995\pm 0.011$ for $\lambda =
49-58$\AA; and $0.959\pm 0.027$ for $\lambda=44-49$\AA. The glitch in
the ratio near 60\AA\ may be a relic of differences between the
positive and negative dispersion directions. We can not exclude an
effect at the percent level given the remaining systematic errors, but
the effect may as well be still smaller as suggested by
\citet{suleimanovetal06}.

\begin{figure}[t]
\vspace*{-0.4mm}
\includegraphics[width=8.8cm]{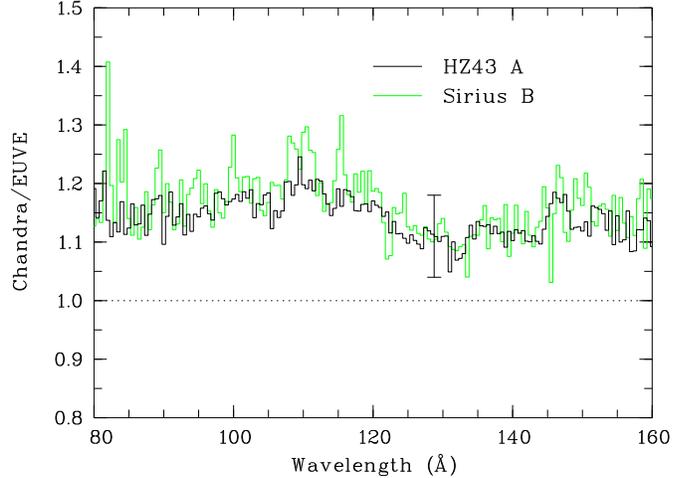}
\caption{Ratio of the soft X-ray fluxes in \ohz\ and \osi\ based on
the \chandra\ LETG+HRC spectra analyzed here and the \euve\ archive
spectra listed in Table~\ref{tab:obslog}. The results are valid for
the calibration using \logg(\ohz)=7.90 and would shift up or down by
the systematic error shown at 128\AA\ for \logg\ lower or higher by
0.1, respectively.}
\label{fig:euve}
\end{figure}

\subsubsection{\osi}

The angular radius for \osi\ in Table~\ref{tab:results} falls between
the values quoted by \citet{barstowetal05} for the HST/STIS G430L and
G750L spectra. At the \textit{Hipparcos} distance of
$2.637\pm0.011$\,pc, the radius of the white dwarf is
$R=\ten{(5.683\pm 0.019)}{8}\,\mathrm{cm}=\ten{(8.165\pm0.028)}{-3}\,\mathrm{R}_\odot$, 
where the error includes the systematic error in the \mbox{LETG+HRC-S}
area correction and is insensitive to \logg\ of \osi\
itself. Combined with the gravitational redshift of
\citet{barstowetal05}, $\upsilon_\mathrm{gr}=80.42\pm
4.83$\,km\,s$^{-1}$, we obtain a mass of $M=1.032\pm0.062\,M_\odot$,
consistent with the astrometric solution of \citet{gatewood78}. The
implied gravity \logg$~=8.627\pm 0.026$ slightly exceeds the
spectroscopic value of \citet{barstowetal05}, but stays within
Holberg's et al. (1998) 1-$\sigma$ error. Conversely, the effective
temperature of $24923\pm 115$\,K (systematic error) falls below
Barstow's spectroscopic value of $25193\pm37$\,K (statistical error)
derived from a Balmer line fit to the HST/STIS spectrum.  The possible
systematic error quoted by \citet{barstowetal05} is 350K and
encompasses our result\footnote{Note that our systematic error of
115\,K does not include the uncertainty in the visual flux of
Sirius~B: a $+0.01$\,mag change (1\% decrease in the adopted 4600\AA\
flux) raises our value of the effective temperature of \osi\ by
14\,K or 0.057\%. It is in the nature of our simultaneous fit that at
the same time the effective temperature of \ohz\ is slightly reduced
(by 0.050\%). The soft X-ray fluxes stay practically unchanged.}. A
spectrum calculated with our code for Barstow's \teff\ and \logg\
adjusted to the same visual HST/STIS flux as above yields a predicted
LETG+HRC spectral flux 14\% higher than our spectrum in
Fig.~\ref{fig:phot}, about twice its log\,$g$-derived systematic error
of 7\%.

\subsubsection{\rxj}

The high quality ($\,\chi^2_\nu=0.87$) of the two-blackbody fit to the
LETG+HRC spectrum of \rxj\ in Figs.~\ref{fig:rxj} and \ref{fig:phot}
demonstrates that the deviation from this model is minute. The model
is normalized to the extinction-corrected optical flux of
\citet{kerkwijkkulkarni01} and provides an almost perfect fit to all
available data.  Blackbody temperature and angular radius of the
hotter component, k$T_\mathrm{spot}=62.8\pm0.4$\,eV and
0.0378\,km\,pc$^{-1}$ (Table~\ref{tab:results}), agree well with
previous determinations
\citep{burwitzetal01,burwitzetal03,brajeromani02,drakeetal02,ponsetal02}.
The temperature k$T_\mathrm{star}=32.3\pm0.7$\,eV of the cooler
component is determined here more accurately than before as a result
of our elimination of some of the deficiencies in the LETGS effective
area calibration at long wavelengths. The angular radius of this
component (Table~\ref{tab:results}) refers to the outer radius of a
blackbody-emitting annulus surrounding the hot spot. The quoted errors
account for the systematic uncertainty in the area correction and
indicate that the main error source for the radii still lies in the
distance uncertainty. The excellence of the two-blackbody fit does not
imply that \rxj\ actually radiates as a blackbody \citep[see, e.g.,
the discussion in][]{kargaltsevetal05}. E.g., a graybody with its
different limb darkening law would yield a different radius than the
disk of uniform brightness. However, whatever the emission properties,
the spectral flux integrated over the surface must nevertheless agree
with that of the blackbody model within the statistical uncertainties
of the X-ray and optical/ultraviolet data. Hence, the blackbody fit
is, in fact, quite restrictive. For the purpose of using \rxj\ as a
soft X-ray calibrator the blackbody assumption is entirely
satisfactory and need be replaced only if future observations reveal
fine structure in the spectra not accounted for by this
approximation. At present, we agree with \citep{brajeromani02} that
there is no evidence for a more complex temperature distribution
beyond the dichotomy of base and spot temperature.
 
\subsection{Calibration of the EUVE SW spectrometer}

It is instructive to compare our photon spectra of \ohz\ and \osi\
with the short-wavelength spectra of these stars in the \euve\ archive
(Table~\ref{tab:obslog}). The mean \euve\ spectra are included in
Fig.~\ref{fig:phot} spectra and are seen to fall below the \chandra\
LETG+HRC spectra in our calibration. Fig.~\ref{fig:euve} shows the
ratios of \chandra\ vs. \euve\ fluxes. They average 1.14 and 1.16 for
\ohz\ and \osi, respectively.  The systematic error of 7\% from
the uncertainty in \logg\ of \ohz\ would shift both numbers up or
down. This comparison suggests that the effective area used for the
short-wavelength section of the \euve\ archive spectra is too high by
a factor of $1.15\pm 0.07$ (systematic error).  A similar
discrepancy between their best-fit models and the EUVE
short-wavelength fluxes was noted previously by \citet{singetal02}.
Interestingly, they found agreement of the models with the EUVE fluxes
at longer wavelengths, a result which has implicitely been adopted in
our approach of determining the hydrogen column density by fitting the
model spectrum to the EUVE long-wavelength flux at 485\AA.  Our
factor of 1.15 at the short wavelengths happens to be close to that
by which the \textit{EUVE} short-wavelength effective area determined
on ground was corrected \emph{upward} after an in-flight calibration
using theoretical spectra of the white dwarfs \ohz, GD71, and GD153
\citep{boydetal94}. In retrospect, that correction seems to have been
unnecessary.

\subsection{\rosat\ PSPC calibration}
\label{sec:pspc}

\begin{figure}[t]
\includegraphics[width=8.8cm]{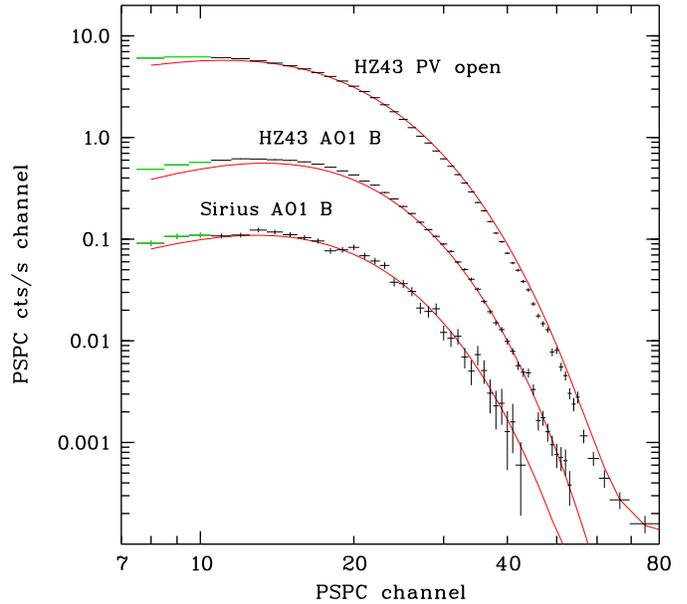}
\caption{\rosat\ PSPC spectra of \ohz\ and \osi\ in either the open
configuration or with boron filter (B). The abscissa is in channels
with one channel corresponding to an apparent photon energy of
10\,eV. The model spectra based on the fits to the LETGS observations
are shown as solid curves, the PSPC observations as data points with
statistical errors. \ohz\ possesses a bremsstrahlung component which
commences at channel 65 and is included in the fit.}
\label{fig:pspc}
\end{figure}

We test the \rosat\ PSPC calibration by folding the photon spectra of
Fig.~\ref{fig:phot} through the PSPC detector response and compare
them with the observed low-resolution PSPC spectra. We consider here
the spectra of \ohz\ and \osi\ which were observed early in the
mission (PV and AO-1 phases) and about two years into the mission
(AO-2), before and after the change in high voltage which affected the
detector response (see Sect.~2.3 and Table~\ref{tab:obslog}).  We have
selected two long observations of \ohz, one in the open configuration
during the PV phase with PSPC-C and one with the boron filter in AO1
with PSPC-B.  \osi\ was observed only once with the boron filter in
AO-1.  We used the appropriate standard detector matrices DRMPSPC-AO1
and DRMPSPC and applied no corrections for gain shifts (see
Sect.~2.3).  The resulting PSPC spectra are shown in
Fig.~\ref{fig:pspc} and reproduce the observations exceedingly
well. The relevant integrated count rates for energy channels 11--60
are listed in Table~\ref{tab:pspc}. The errors given for the predicted
count rates refer to the systematic uncertainty in our photon spectra
and \mbox{LETG+HRC-S} area correction discussed above.  The observed
count rate of \ohz\ in the open configuration agrees perfectly with
the prediction.  Hence, we can safely conclude that the calibration of
the PSPC is correct and the suggestions of a miscalibration by
\citet{napiwotzkietal93}, \citet{jordanetal94}, and
\citet{wolffetal95,wolffetal96} are unfounded.  Close inspection
reveals some small discrepancies between prediction and observation
which find ready explanations in a variety of small necessary
corrections. In the uppermost spectrum in Fig.~\ref{fig:pspc}, the
HZ43 observation in the open mode, the observed spectrum is slightly
shifted to lower energy channels. This effect is caused by the gain
depression by some 4\% in the small central focal spot
\citep{snowdenetal01}.  The observational `wobble' mode moves the
source forth and back across this central depression and causes a mean
gain reduction of about 1.5\% which accounts for the observed shift.
Of a different origin is the slight excess in count rate for the
observations with boron filter (lower two spectra in
Fig.~\ref{fig:pspc}). A probable explanation is provided by an actual
thickness of the boron window about 4\% lower than nominal.

\begin{table}[t]
\caption{Observed PSPC count rates in channels 11--60 and
LETGS-predicted count rates based on the photon spectra in
Fig.~\ref{fig:phot}. `O' and `B' in column 3 refer to observations in the
open configuration and with the boron filter, respectively.}
\label{tab:pspc}
\begin{tabular}{lccccc} 
\hline \hline \noalign{\smallskip}

             &        &        &        & \multicolumn{2}{c}{Count rate} \\ 
Object       &  Date  & Filter & Exp. &  observed & \hspace{-2mm}predicted                  \\ 
             &  Start &        & (ks)  &  (cts/s)  & \hspace{-2mm} (cts/s) \\   
                               
\noalign{\smallskip} \hline
\noalign{\smallskip}
\ohz         & 900621 & O   & \hspace{-2mm} 21.5 &\hspace{-2mm} $66.08\pm 0.30$ &\hspace{-2mm}$65.38\pm 2.70$\\
             & 910619 & B  & \hspace{-2mm} 21.6 & $ 8.04\pm 0.03$ & $7.35\pm0.28$\\[0.5ex]
\osi         & 910315 & B  &   3.3  & $ 1.43\pm 0.03$ & $1.38\pm 0.05$\\[0.5ex]
\rxj         & 921111 & O   &   6.3  & $ 3.51\pm 0.03$ & $3.33\pm 0.17$\\[0.5ex]
\noalign{\smallskip} \hline      
			         
\end{tabular}
\end{table}

The remaining small calibration uncertainties of the PSPC and the
calibration of the high resolution imager (HRI) will be addressed
elsewhere.  All these effects are at the few percent level and provide
no basis for the allegation of a miscalibration. In summary, the
internal consistency of our results suggests that the PSPC yields a
correct measurement of the soft X-ray flux and the ground calibration
of the detector at soft X-ray energies is reliable to better than
10\%, as concluded also by \citet{snowdenetal95}. Given the large
number of observations of all types of celestial sources, both in the
All-Sky Survey and in pointed observations, we judge this a reassuring
statement of general interest. One can only speculate on the basis of
the reports about a major miscalibration of the PSPC. The use of model
spectra which provided perfect fits to observations in the optical and
ultraviolet but lacked flux in the soft X-ray regime is a possibility.

In passing, we confirm that \ohz\ possesses a bremsstrahlung component
with a temperature \mbox{of 0.6 keV} and an emission measure of
$\ten{3}{52}$\,\cmcub, which is seen to commence near channel 65 in
Fig.~\ref{fig:pspc}. It was previously reported by
\citet{odwyeretal03} and probably originates from the secondary star.
We have checked the zeroth order images of all individual LETGS
observations for a signal at the position of the secondary (separation
3\,arcsec, position angle $280^\circ$), but the point spread function
of the white dwarf image still contributes a flux about an order of
magnitude larger than expected for the secondary even if the position
of the secondary is favorably located between the diffraction spikes.
Hence, an origin of the hard X-rays from the secondary is plausible
but remains unproven.

\section{Conclusions}

We have established the neutron star \rxj\ and the hot white dwarfs
\ohz\ and \osi\ as standard stars in the soft X-ray regime by fitting
the LETG+HRC spectra of all three stars simultaneously with the best
available model spectra for the white dwarfs and a two-blackbody model
for the neutron star. Our method ties the soft X-ray spectra of the
two white dwarfs to that of the neutron star \rxj\ and thereby to be
better calibrated effective areas of the \chandra\ LETG+HRC
spectrometer shortwards of 40\AA. The spectrum of \ohz\ (and
indirectly that of \osi) is thus fixed at soft X-ray \emph{and} at
optical wavelengths, a procedure which avoids the wagging-tail problem
of previous calibration attempts using white dwarfs. Our second major
result is the self-consistently determined correction to the
LETG+HRC-S effective areas longwards of 44\AA. We find that the areas
for both dispersion directions need a maximum correction of about
$-23$\% at 65\AA\ and are correct to within $\sim10$\% at wavelengths
exceeding 80\AA. These corrections are valid relative to the
short-wavelength effective areas. The largest remaining uncertainty,
which is not addressed here, lies in the absolute calibration of the
LETG+HRC-S at short wavelengths.  Our photospheric parameters of the
three stars are in good agreement with those derived from independent
optical/ultraviolet and soft X-ray studies and, in part, improve on
them. The soft X-ray flux of \ohz\ can be produced by a range of
\teff-\logg\ combinations and is better defined than either of these
two parameters.

We find that the calibration of the \euve\ short-wavelength
spectrometer differs from that of the \chandra\ LETG+HRC-S by about
$15\pm 7$\% in the sense that the original ground calibration seems to
have been more adequate than the adopted in-flight calibration.

We demonstrate that the ground calibration of the \rosat\ PSPC is
correct to within a few percent and that previous reports of a major
miscalibration are unfounded. The observed PSPC spectra of our three
stars agree within a few percent with the LETGS-based soft X-ray
spectra folded through the PSPC response. This internal consistency,
in turn, supports our adjustment of the \mbox{LETG+HRC-S} calibration
at long wavelengths and the basic correctness of the nominal
calibration at short wavelengths.

\begin{acknowledgements}
This research was supported by the DLR under grant 50\,OR\,0201
(TR). We thank the referee Jay Holberg for helpful questions and
comments, our colleagues from the \rosat\ team and numerous
colleagues from the general community for discussions, advice, and
support, Frank Haberl for contributions in the early phase of this
project, and Michael Freyberg for information on the calibration of
the \rosat\ PSPC. KB is indebted to Stefan Dreizler, Boris G\"ansicke,
Frederik Hessman, Klaus Reinsch, and Sonja Schuh for many discussions,
and to Joachim Tr\"umper for helpful comments and criticism.
\end{acknowledgements}

\bibliographystyle{aa}

\end{document}